\documentstyle[12pt]{article}

\catcode`\@=11
\def\numberbysection{\@addtoreset{equation}{section}
        \def\theequation{\thesection.\arabic{equation}}}
\numberbysection

\input{tcilatex}

\begin{document}

\newlength{\lno} \lno1.5cm \newlength{\len} \len=\textwidth%
\addtolength{\len}{-\lno}

\setcounter{page}{0}

\baselineskip7mm \renewcommand{\thefootnote}{\fnsymbol{footnote}} \newpage %
\setcounter{page}{0}

\begin{titlepage}     
\vspace{0.5cm}
\begin{center}
{\Large\bf osp{$(1|2)$} Off-shell Bethe Ansatz Equations}\\
\vspace{1cm}
{\large A. Lima-Santos }\footnote{e-mail: dals@power.ufscar.br} \\
\vspace{1cm}
{\large \em Universidade Federal de S\~ao Carlos, Departamento de F\'{\i}sica \\
Caixa Postal 676, CEP 13569-905~~S\~ao Carlos, Brasil}\\
\end{center}
\vspace{1.2cm}

\begin{abstract}
The semiclassical limit of the algebraic quantum inverse scattering method
is used to solve the theory of the Gaudin model. Via Off-shell Bethe ansatz
equations of an integrable representation of the graded $osp(1|2)$ vertex
model we find the spectrum of the $N-1$ independents Hamiltonians of Gaudin.
Integral representations of the $N$-point correlators are presented as
solutions of the Knizhnik-Zamolodchikov equation. These results \ are
extended for highest representations of the $osp(1|2)$ Gaudin algebra.
\end{abstract}
\vfill
\begin{center}
\small{\today}
\end{center}
\end{titlepage}

\baselineskip6mm

\newpage{}

\section{Introduction}

In integrable models of statistical mechanics \cite{Baxter}, an important
object is the ${\cal R}$-matrix ${\cal R}(\lambda )$, where $\lambda $ is
the spectral parameter. It acts on the tensor product $V^{1}\otimes V^{2}$
for a given vector space $V$ and it is the solution of the Yang-Baxter (%
{\small YB}) equation 
\begin{equation}
{\cal R}_{12}(\lambda ){\cal R}_{13}(\lambda +\mu ){\cal R}_{23}(\mu )={\cal %
R}_{23}(\mu ){\cal R}_{13}(\lambda +\mu ){\cal R}_{12}(\lambda )
\label{int.1}
\end{equation}
in $V^{1}\otimes V^{2}\otimes V^{3}$, where ${\cal R}_{12}={\cal R}\otimes 
{\cal I}$, ${\cal R}_{23}={\cal I}\otimes {\cal R}$, etc. and ${\cal I}$ \
is the identity matrix. If ${\cal R}$ depends on a Planck-type parameter $%
\eta $ so that ${\cal R}(\lambda ,\eta )=1+\eta \ r(\lambda )+{\rm o}(\eta
^{2})$, as $\eta \rightarrow 0$, then the ``classical $r$-matrix'' obeys the
classical {\small YB} equation 
\begin{equation}
\lbrack r_{12}(\lambda ),r_{13}(\lambda +\mu )+r_{23}(\mu )]+[r_{13}(\lambda
+\mu ),r_{23}(\mu )]=0  \label{int.2}
\end{equation}

Nondegenerate solutions of (\ref{int.2}) in the tensor product of two copies
of simple Lie algebra {\rm g} , $r_{ij}(\lambda )\in {\rm g}_{i}\otimes {\rm %
g}_{j}$ , $i,j=1,2,3$, were classified by Belavin and Drinfeld \cite{BD}.

The classical {\small YB} equation interplays with conformal field theory in
the following way: In the skew-symmetric case $r_{ji}(-\lambda
)+r_{ij}(\lambda )=0$, it is the compatibility condition for the system of
linear differential equations 
\begin{equation}
\kappa \frac{\partial }{\partial z_{i}}\Psi (z_{1},...,z_{N})=\sum_{j\neq
i}r_{ij}(z_{i}-z_{j})\Psi (z_{1},...,z_{N})  \label{int.3}
\end{equation}
in $N$ complex variables $z_{1},...,z_{N}$ for vector-valued functions $\Psi 
$ with values in the tensor space $V=V^{1}\otimes \cdots \otimes V^{N}$ and $%
\kappa $ is a coupling constant.

In the rational case \cite{BD}, very simple skew-symmetric solutions are
known: $r(\lambda )={\rm C}_{2}/\lambda $, where ${\rm C}_{2}\in {\rm g}%
\otimes {\rm g}$ is a symmetric invariant tensor of a finite dimensional Lie
algebra ${\rm g}$ acting on a representation space $V$. The corresponding
system of linear differential equations is the Knizhnik-Zamolodchikov (%
{\small KZ}) equation for the conformal blocks of the
Wess-Zumino-Novikov-Witten ({\small WZNW}) model of conformal theory on the
sphere \cite{KZ}. Therefore, one can consider a {\small KZ} equation for
each solution of the classical {\small YB} equation.

The algebraic Bethe ansatz \cite{FT} is the powerful method in the analysis
of integrable models. Besides describing the spectra of quantum integrable
systems, the Bethe ansatz also is used to find reasonably efficient
expressions for the correlators \cite{KBI}. Various representations of
correlators in these models were found by Korepin \cite{KO}, using this
method.

Recently, Babujian and Flume \cite{BAF} developed a method which reveals a
link to the algebraic Bethe ansatz for the theory of the Gaudin model. Their
method was confirmed by another different approach presented in the work of
Feigin, Frenkel and Reshetikhin \cite{FFR}. In the Babujian-Flume method the
wave vectors of the Bethe ansatz equation for inhomogeneous lattice model
render in the semiclassical limit solutions of the {\small KZ} equation for
the case of simple Lie algebras of higher rank. More precisely, the
algebraic quantum inverse scattering method permits us write the following
equation 
\begin{equation}
t(\lambda |z)\Phi (\lambda _{1,\cdots ,}\lambda _{p})=\Lambda (\lambda
,\lambda _{1},\cdots ,\lambda _{p}|z)\Phi (\lambda _{1},\cdots ,\lambda
_{p})-\sum_{\alpha =1}^{p}\frac{{\cal F}_{\alpha }\Phi ^{\alpha }}{\lambda
-\lambda _{\alpha }}  \label{int.4}
\end{equation}
Here $t(\lambda |z)$ denotes the transfer matrix of the rational vertex
model in an inhomogeneous lattice acting on an $N$-fold tensor product of $%
SU(2)$ representation spaces. $\Phi ^{\alpha }=\Phi (\lambda _{1},\cdots
\lambda _{\alpha -1},\lambda ,\lambda _{\alpha +1},...,\lambda _{p})$. $%
{\cal F}_{\alpha }(\lambda _{1},\cdots ,\lambda _{p}|z)$ and $\Lambda
(\lambda ,\lambda _{1},\cdots ,\lambda _{p}|z)$ are $c$ numbers. The
vanishing of the so-called unwanted terms, ${\cal F}_{\alpha }=0$, is
enforced in the usual procedure of the algebraic Bethe ansatz by choosing
the parameters $\lambda _{1},...,\lambda _{p}$. In this case the wave vector 
$\Phi (\lambda _{1},\cdots ,\lambda _{p})$ becomes an eigenvector of the
transfer matrix with eigenvalue $\Lambda (\lambda ,\lambda _{1},\cdots
,\lambda _{p}|z)$. If we keep all unwanted terms, i.e. ${\cal F}_{\alpha
}\neq 0$, then the wave vector $\Phi $ in general satisfies the equation (%
\ref{int.4}), named in \cite{B} as off-shell Bethe ansatz equation ({\small %
OSBAE}). There is a neat relationship between the wave vector satisfying the 
{\small OSBAE} (\ref{int.4}) and the vector-valued solutions of the {\small %
KZ} equation (\ref{int.3}): The general vector valued solution of the 
{\small KZ} equation for an arbitrary simple Lie algebra was found by
Schechtman and Varchenko \cite{SV}. It can be represented as a multiple
contour integral

\begin{equation}
\Psi (z_{1},\ldots ,z_{N})=\oint \cdots \oint {\cal X}(\lambda
_{1},...,\lambda _{p}|z)\phi (\lambda _{1},...,\lambda _{p}|z)d\lambda
_{1}\cdots d\lambda _{p}  \label{int.5}
\end{equation}
The complex variables $z_{1},...,z_{N}$ of (\ref{int.5}) are related with
the disorder parameters of the {\small OSBAE} . The vector valued function $%
\phi (\lambda _{1},...,\lambda _{p}|z)$ is the semiclassical limit of the
wave vector $\Phi (\lambda _{1},...,\lambda _{p}|z)$. In fact, it is the
Bethe wave vector for Gaudin magnets \cite{GA}, but off mass shell. The
scalar function ${\cal X}(\lambda _{1},...,\lambda _{p}|z)$ is constructed
from the semiclassical limit of the $\Lambda (\lambda =z_{k};\lambda
_{1},...,\lambda _{p}|z)$ and ${\cal F}_{\alpha }(\lambda _{1},\cdots
,\lambda _{p}|z)$. This representation of the $N$-point correlation function
shows a deep connection between the inhomogeneous vertex models and the 
{\small WZNW }theory.

In this paper we apply the Babujian-Flume ideas for $osp(1|2)$ rational
solution of the graded version of the {\small YB} equation (\ref{int.1}).

The paper is organized as follows. In Section $2$ we present the algebraic
structure of the $osp(1|2)$ vertex model. The inhomogeneous Bethe ansatz is
read of from the homogeneous case previously known \cite{LI}. We derive the
Off-shell Bethe ansatz equations for the fundamental representation of the $%
osp(1|2)$ algebra. In Section $3$ , taking into account the semiclassical
limit ($\eta \rightarrow 0$) of the results presented in the Section $2$, we
describe the algebraic structure of the corresponding Gaudin model. Using
the $osp(1|2)$ Gaudin algebra, these results are extended for highest
representations. In Section $4$, data of the Gaudin off-shell Bethe ansatz
equations are used to construct solutions of the {\small KZ} equation.
Conclusions are reserved for Section $5$.

\section{Structure of the ${\bf osp(1|2)}$ Vertex Model}

We recall that the $osp(1|2)$ algebra is the simplest superalgebra and it
can be viewed as the graded version of $sl(2)$. It has three even (bosonic)
generators $H,$\ $X^{\pm }$ generating a Lie subalgebra $sl(2)$ and two odd
(fermionic) generators $V^{\pm }$ , whose non-vanishing commutation
relations in the Cartan-Weyl basis reads as 
\begin{eqnarray}
\lbrack H,X^{\pm }] &=&\pm X^{\pm },\quad \lbrack X^{+},X^{-}]=2H  \nonumber
\\
\lbrack H,V^{\pm }] &=&\pm \frac{1}{2}V^{\pm },\quad \lbrack X^{\pm },V^{\mp
}]=V^{\pm },\quad \lbrack X^{\pm },V^{\pm }]=0  \nonumber \\
\{V^{\pm },V^{\pm }\} &=&\pm \frac{1}{2}X^{\pm },\quad \{V^{+},V^{-}\}=-%
\frac{1}{2}H  \label{str.1}
\end{eqnarray}
The quadratic Casimir operator is 
\begin{equation}
C_{2}=H^{2}+\frac{1}{2}\{X^{+},X^{-}\}+[V^{+},V^{-}]  \label{str.2}
\end{equation}
where $\{\cdot \ ,\cdot \}$ denotes the anticommutator and $[\cdot \ ,\cdot
] $ the commutator.

The irreducible finite-dimensional representations $\rho _{j}$ with the
highest weight vector are parametrized by half-integer $s=j/2$ or by the
integer $j\in N$. Their dimension is {\rm dim}$(\rho _{j})=2j+1$ and the
corresponding value of $C_{2}$ is $j(j+1)/4=s(s+1/2)$, $s=0,1/2,1,3/2,...$

The representation corresponding to $s=0$ is the trivial one-dimensional
representation. The $s\geq 1/2$ representation contains two isospin
multiplets which belong to isospin $s$ and $s-1/2$, denoted by $\left|
s,s,m\right\rangle $ and $\left| s,s-1/2,m\right\rangle $, respectively. The
first quantum number characterizes the representation and the second and
third quantum numbers give the isospin and its third component. After a
convenient normalization of the states , a given $s$-representation is
defined by 
\begin{eqnarray}
H\left| s,s,m\right\rangle &=&m\left| s,s,m\right\rangle ,  \nonumber \\
X^{\pm }\left| s,s,m\right\rangle &=&\sqrt{(s\mp m)(s\pm m+1)}\left|
s,s,m\pm 1\right\rangle  \nonumber \\
V^{\pm }\left| s,s,m\right\rangle &=&\pm \frac{1}{2}\sqrt{(s\mp m)}\left|
s,s-1/2,m\pm 1/2\right\rangle  \nonumber \\
&&  \nonumber \\
\quad H\left| s,s-1/2,m\right\rangle &=&m\left| s,s-1/2,m\right\rangle 
\nonumber \\
X^{\pm }\left| s,s-1/2,m\right\rangle &=&\sqrt{(s-1/2\mp m)(s-1/2\pm m+1)}%
\left| s,s-1/2,m\pm 1\right\rangle  \nonumber \\
V^{\pm }\left| s,s-1/2,m\right\rangle &=&\pm \frac{1}{2}\sqrt{(s-1/2\pm m+1)}%
\left| s,s,m\pm 1/2\right\rangle  \label{str.3}
\end{eqnarray}
The fundamental representation has $s=1/2$ and is given by 
\begin{eqnarray}
H &=&\frac{1}{2}\left( 
\begin{array}{lll}
1 & 0 & \ \ 0 \\ 
0 & 0 & \ \ 0 \\ 
0 & 0 & -{}1
\end{array}
\right) ,\ \ X^{+}=\left( 
\begin{array}{lll}
0 & 0 & 1 \\ 
0 & 0 & 0 \\ 
0 & 0 & 0
\end{array}
\right) ,\ X^{-}=\left( 
\begin{array}{lll}
0 & 0 & 0 \\ 
0 & 0 & 0 \\ 
1 & 0 & 0
\end{array}
\right)  \nonumber \\
V^{+} &=&\frac{1}{2}\left( 
\begin{array}{lll}
0 & 1 & 0 \\ 
0 & 0 & 1 \\ 
0 & 0 & 0
\end{array}
\right) ,\ V^{-}=\frac{1}{2}\left( 
\begin{array}{lll}
\ \ 0 & 0 & 0 \\ 
-1 & 0 & 0 \\ 
\ \ 0 & 1 & 0
\end{array}
\right)  \label{str.4}
\end{eqnarray}

In (\ref{str.4}) the basis is $\left| \frac{1}{2},\frac{1}{2},\frac{1}{2}%
\right\rangle ,\left| \frac{1}{2},0,0\right\rangle ,\left| \frac{1}{2},\frac{%
1}{2},-\frac{1}{2}\right\rangle $. The first and third vectors will be
considered as even and the second as odd, {\it i.e}., our grading is {\small %
BFB}.

In the $j$-representation, the odd part has the form \cite{Ku1}: 
\begin{equation}
V^{+}=\left( 
\begin{array}{ccccc}
0 & V_{j-1} & 0 & \cdots  & 0 \\ 
0 & 0 & V_{j-2} & \ddots  & \vdots  \\ 
\vdots  & \ddots  & \ddots  & \ddots  & 0 \\ 
0 & \cdots  & 0 & 0 & V_{-j} \\ 
0 & \cdots  & 0 & 0 & 0
\end{array}
\right) ,\quad V^{-}=\left( 
\begin{array}{ccccc}
0 & 0 & \cdots  & 0 & 0 \\ 
W_{j} & 0 & \ddots  & \vdots  & \vdots  \\ 
0 & W_{j-1} & \ddots  & 0 & 0 \\ 
\vdots  & \ddots  & \ddots  & 0 & 0 \\ 
0 & \cdots  & 0 & W_{-j+1} & 0
\end{array}
\right)   \label{str.5}
\end{equation}
where 
\begin{eqnarray}
(V_{j-1},V_{j-2},V_{j-3},...,V_{-j}) &=&\frac{1}{2}(\sqrt{j},\sqrt{1},\sqrt{%
j-1},\sqrt{2},...,\sqrt{1},\sqrt{j}),  \nonumber \\
(W_{j},W_{j-1},W_{j-2},...,W_{-j+1}) &=&\frac{1}{2}(-\sqrt{j},\sqrt{1},-%
\sqrt{j-1},\sqrt{2,}...,-\sqrt{1},\sqrt{j}).  \label{str.6}
\end{eqnarray}
For the even part we can see from (\ref{str.3}) that $H$ is diagonal and
always has eigenvalue $0$ due to isospin integer: 
\begin{equation}
H=\frac{1}{2}{\rm diag}(j,j-1,...,1,0,-1,...,-j).  \label{str.7}
\end{equation}
Moreover, $X^{\pm }$ are given by the $sl(2)$ composition which results in a
clear relation with the odd part: $X^{\pm }=\pm 4(V^{\pm })^{2}$.

\subsection{Graded Quantum Inverse Scattering Method}

Consider $V=V_{0}\oplus V_{1}$ a $Z_{2}$-graded vector space where $0$ and $%
1 $ denote the even and odd parts respectively. The components of a linear
operator $A\stackrel{s}{\otimes }B$ in the graded tensor product space $V%
\stackrel{s}{\otimes }V$ result in matrix elements of the form 
\begin{equation}
(A\stackrel{s}{\otimes }B)_{\alpha \beta }^{\gamma \delta }=(-)^{p(\beta
)(p(\alpha )+p(\gamma ))}\ A_{\alpha \gamma }B_{\beta \delta }  \label{gra.1}
\end{equation}
and the action of the permutation operator ${\cal P}$ on the vector $\left|
\alpha \right\rangle \stackrel{s}{\otimes }\left| \beta \right\rangle \in V%
\stackrel{s}{\otimes }V$ is given by 
\begin{equation}
{\cal P}\ \left| \alpha \right\rangle \stackrel{s}{\otimes }\left| \beta
\right\rangle =(-)^{p(\alpha )p(\beta )}\left| \beta \right\rangle \stackrel{%
s}{\otimes }\left| \alpha \right\rangle \Longrightarrow ({\cal P})_{\alpha
\beta }^{\gamma \delta }=(-)^{p(\alpha )p(\beta )}\delta _{\alpha \beta }\
\delta _{\gamma \delta }  \label{gra.2}
\end{equation}
where $p(\alpha )=1\ (0)$ if $\left| \alpha \right\rangle $ is an odd (even)
element.

Besides ${\cal R}$ we have to consider matrices $R={\cal PR}$ which satisfy 
\begin{equation}
R_{12}(\lambda )R_{23}(\lambda +\mu )R_{12}(\mu )=R_{23}(\mu )R_{12}(\lambda
+\mu )R_{23}(\lambda ).  \label{gra.2a}
\end{equation}

The rational solution of the graded {\small YB} equation for the fundamental
representation of $osp(1|2)$ algebra was found by Kulish in \cite{Ku2}. It
has the form 
\begin{equation}
R(\lambda ,\eta )=\left( 
\begin{array}{lllllllll}
x_{1} & 0 & \ 0 & 0 & \ 0 & 0 & \ 0 & 0 & 0 \\ 
0 & x_{5} & \ 0 & x_{2} & \ 0 & 0 & \ 0 & 0 & 0 \\ 
0 & 0 & \ x_{7} & 0 & \ y_{6} & 0 & \ x_{3} & 0 & 0 \\ 
0 & x_{2} & \ 0 & x_{5} & \ 0 & 0 & 0 & 0 & 0 \\ 
0 & 0 & -y_{6} & 0 & -x_{4} & 0 & -x_{6} & 0 & 0 \\ 
0 & 0 & \ 0 & 0 & \ 0 & x_{5} & \ 0 & x_{2} & 0 \\ 
0 & 0 & \ x_{3} & 0 & \ x_{6} & 0 & \ x_{7} & 0 & 0 \\ 
0 & 0 & \ 0 & 0 & \ 0 & x_{2} & \ 0 & x_{5} & 0 \\ 
0 & 0 & \ 0 & 0 & \ 0 & 0 & \ 0 & 0 & x_{1}
\end{array}
\right)  \label{gra.3}
\end{equation}
where 
\begin{eqnarray}
x_{1}(\lambda ,\eta ) &=&\lambda +\eta ,\quad x_{2}(\lambda ,\eta )=\epsilon
_{1}\lambda ,\quad x_{3}(\lambda ,\eta )=\lambda -\frac{\lambda \eta }{%
\lambda +\frac{3}{2}\eta }  \nonumber \\
x_{4}(\lambda ,\eta ) &=&\lambda -\eta +\frac{\lambda \eta }{\lambda +\frac{3%
}{2}\eta },\quad x_{5}(\lambda ,\eta )=\eta ,\quad x_{6}(\lambda ,\eta
)=-\epsilon _{2}\frac{\lambda \eta }{\lambda +\frac{3}{2}\eta }  \nonumber \\
x_{7}(\lambda ,\eta ) &=&\eta +\frac{\lambda \eta }{\lambda +\frac{3}{2}\eta 
},\qquad y_{6}(\lambda ,\eta )=-x_{6}(\lambda ,\eta ),\quad  \label{gra.4}
\end{eqnarray}
where $\epsilon _{i}=\pm 1,$ $i=1,2$. Here we have assumed that the grading
of threefold space is $p(1)=p(3)=0$ and $p(2)=1$ and we will choose the
solution of (\ref{gra.4}) for which $\epsilon _{1}=\epsilon _{2}=1$.

Let us consider the inhomogeneous vertex model, where to each vertex we
associate two parameters: the global spectral parameter $\lambda $ and the
disorder parameter $z$. In this case, the vertex weight matrix ${\cal R}$
depends on $\lambda -z$ and consequently the monodromy matrix will be a
function of the disorder parameters $z_{i}$. From now on we use a compact
notation for the arguments with the shifted spectral parameter: $(\lambda
|z)=(\lambda -z_{1},...,\lambda -z_{N})$.

The graded quantum inverse scattering method is characterized by the
monodromy matrix $T(\lambda |z)$ satisfying the equation 
\begin{equation}
R(\lambda -\mu )\left[ T(\lambda |z)\stackrel{s}{\otimes }T(\mu |z)\right] =%
\left[ T(\mu |z)\stackrel{s}{\otimes }T(\lambda |z)\right] R(\lambda -\mu ),
\label{gra.5}
\end{equation}
whose consistency is guaranteed by the graded version of the {\small YB}
equation (\ref{gra.2a}). $T(\lambda |z)$ is a matrix in the space $V$ (the
auxiliary space) whose matrix elements are operators on the states of the
quantum system (the quantum space, which will also be the space $V$). The
monodromy operator $T(\lambda |z)$ is defined as an ordered product of local
operators ${\cal L}_{n}$ (Lax operator), on all sites of the lattice: 
\begin{equation}
T(\lambda |z)={\cal L}_{N}(\lambda |z){\cal L}_{N-1}(\lambda |z)\cdots {\cal %
L}_{1}(\lambda |z).  \label{gra.6}
\end{equation}
The Lax operator on the $n^{th}$ quantum space is given the graded
permutation of (\ref{gra.3}): 
\begin{eqnarray}
{\cal L}_{n}(\lambda |z) &=&\left( 
\begin{array}{lllllllll}
x_{1} & 0 & 0 & 0 & \ \ 0 & 0 & \ \ 0 & 0 & 0 \\ 
0 & x_{2} & 0 & x_{5} & \ \ 0 & 0 & \ \ 0 & 0 & 0 \\ 
0 & 0 & x_{3} & 0 & x_{6} & 0 & \ x_{7} & 0 & 0 \\ 
0 & x_{5} & 0 & x_{2} & \ \ 0 & 0 & \ \ 0 & 0 & 0 \\ 
0 & 0 & y_{6} & 0 & x_{4} & 0 & x_{6} & 0 & 0 \\ 
0 & 0 & 0 & 0 & \ \ 0 & x_{2} & \ \ 0 & x_{5} & 0 \\ 
0 & 0 & x_{7} & 0 & \ y_{6} & 0 & \ x_{3} & 0 & 0 \\ 
0 & 0 & 0 & 0 & \ \ 0 & x_{5} & \ \ 0 & x_{2} & 0 \\ 
0 & 0 & 0 & 0 & \ \ 0 & 0 & \ \ 0 & 0 & x_{1}
\end{array}
\right)  \nonumber \\
&=&\left( 
\begin{array}{lll}
L_{11}^{(n)}(\lambda |z) & L_{12}^{(n)}(\lambda |z) & L_{13}^{(n)}(\lambda
|z) \\ 
L_{21}^{(n)}(\lambda |z) & L_{22}^{(n)}(\lambda |z) & L_{23}^{(n)}(\lambda
|z) \\ 
L_{31}^{(n)}(\lambda |z) & L_{32}^{(n)}(\lambda |z) & L_{33}^{(n)}(\lambda
|z)
\end{array}
\right)  \label{gra.7}
\end{eqnarray}
Note that $L_{\alpha \beta }^{(n)}(\lambda ),\ \alpha ,\beta =1,2,3$ are $3$
by $3$ matrices acting on the $n^{th}$ site of the lattice. It means that
the monodromy matrix has the form 
\begin{equation}
T(\lambda |z)=\left( 
\begin{array}{lll}
A_{1}(\lambda |z) & B_{1}(\lambda |z) & B_{2}(\lambda |z) \\ 
C_{1}(\lambda |z) & A_{2}(\lambda |z) & B_{3}(\lambda |z) \\ 
C_{2}(\lambda |z) & C_{3}(\lambda |z) & A_{3}(\lambda |z)
\end{array}
\right)  \label{gra.8}
\end{equation}
where 
\begin{eqnarray}
T_{ij}(\lambda |z) &=&\sum_{k_{1},...,k_{N-1}=1}^{3}L_{ik_{1}}^{(N)}(\lambda
|z)\stackrel{s}{\otimes }L_{k_{1}k_{2}}^{(N-1)}(\lambda |z)\stackrel{s}{%
\otimes }\cdots \stackrel{s}{\otimes }L_{k_{N-1}j}^{(1)}(\lambda |z) 
\nonumber \\
i,j &=&1,2,3.  \label{gra.9}
\end{eqnarray}

The vector in the quantum space of the monodromy matrix $T(\lambda |z)$ that
is annihilated by the operators $T_{ij}(\lambda |z)$, $i>j$ ($C_{i}(\lambda
|z)$ operators, $i=1,2,3$) and it is also an eigenvector for the operators $%
T_{ii}(\lambda |z)$ ( $A_{i}(\lambda |z)$ operators, $i=1,2,3$) is called a
highest vector of the monodromy matrix $T(\lambda |z)$.

The transfer matrix $\tau (\lambda |z)$ of the corresponding integrable spin
model is given by the supertrace of the monodromy matrix in the space $V$ 
\begin{equation}
\tau (\lambda |z)=\sum_{i=1}^{3}(-1)^{p(i)}\ T_{ii}(\lambda
|z)=A_{1}(\lambda |z)-A_{2}(\lambda |z)+A_{3}(\lambda |z).  \label{gra.10}
\end{equation}

A detailed exposition of the graded quantum inverse scattering method can be
found in the references \cite{Ku3},\cite{Ku4} and \cite{EK}.

\subsection{Inhomogeneous Bethe Ansatz}

Algebraic Bethe ansatz solution for the inhomogeneous $osp(1|2)$ vertex
model can be obtained from the homogeneous case. The only modification one
have to do is local shifting of the spectral parameter $\lambda \rightarrow
\lambda -z_{i}$. The algebraic Bethe ansatz solution of the homogeneous
rational $osp(1|2)$ vertex model was obtained by Martins in \cite{MJ}.
However, we will consider the rational limit of the trigonometric case
presented in \cite{LI}.

In the sequence we will work with some functions which will be now defined:

\begin{eqnarray}
z(\lambda ) &=&\frac{x_{1}(\lambda )}{x_{2}(\lambda )}=\frac{\lambda +\eta }{%
\lambda },\quad y(\lambda )=\frac{x_{3}(\lambda )}{y_{6}(\lambda )}=\frac{%
2\lambda +\eta }{2\eta },\quad  \nonumber \\
\omega (\lambda ) &=&-\frac{x_{1}(\lambda )x_{3}(\lambda )}{x_{4}(\lambda
)x_{3}(\lambda )-x_{6}(\lambda )y_{6}(\lambda )}=-\frac{2\lambda +\eta }{%
2\lambda -\eta }  \nonumber \\
{\cal Z}(\lambda _{k}-\lambda _{j}) &=&\left\{ 
\begin{array}{c}
z(\lambda _{k}-\lambda _{j})\qquad \qquad \quad \quad {\rm if}\quad k>j \\ 
z(\lambda _{k}-\lambda _{j})\omega (\lambda _{j}-\lambda _{k})\quad \ {\rm if%
}\quad k<j
\end{array}
\right.  \label{inh.1}
\end{eqnarray}
We start defining the highest vector of the monodromy matrix $T(\lambda |z)$
in a lattice of $N$ sites as the even (bosonic) completely unoccupied state 
\begin{equation}
\left| 0\right\rangle =\otimes _{a=1}^{N}\left( 
\begin{array}{c}
1 \\ 
0 \\ 
0
\end{array}
\right) _{a}  \label{inh.2}
\end{equation}
Using (\ref{gra.9}) we can compute the action of the monodromy matrix
entries on this state 
\begin{eqnarray}
A_{i}(\lambda |z)\left| 0\right\rangle &=&X_{i}(\lambda |z)\left|
0\right\rangle ,\quad C_{i}(\lambda |z)\left| 0\right\rangle =0,\quad
B_{i}(\lambda |z)\left| 0\right\rangle \neq \left\{ 0,\left| 0\right\rangle
\right\}  \nonumber \\
X_{i}(\lambda |z) &=&\prod_{a=1}^{N}x_{i}(\lambda -z_{a}),\qquad i=1,2,3
\label{inh.3}
\end{eqnarray}

Unlike to the simple Lie algebras cases (\ref{int.4}), the off-shell Bethe
ansatz equations for the $osp(1|2)$ algebra present in a more complicated
form

\begin{equation}
\tau (\lambda |z)\Psi _{n}(\lambda _{1},...,\lambda _{n})=\Lambda \Psi
_{n}(\lambda _{1},...,\lambda _{n})-\sum_{j=1}^{n}{\cal F}_{j}^{(n)}\Psi
_{(n)}^{j}+\sum_{j=2}^{n}\sum_{l=1}^{j-1}{\cal F}_{lj}^{(n)}\Psi _{(n)}^{jl}
\label{inh.4}
\end{equation}
where the Bethe vectors are defined as normal ordered states $\Psi
_{n}(\lambda _{1},\cdots ,\lambda _{n})$ which can be written with aid of a
recurrence formula \cite{TA}: 
\begin{eqnarray}
&&\left. \Psi _{n}(\lambda _{1},...,\lambda _{n})=B_{1}(\lambda _{1})\Psi
_{n-1}(\lambda _{2},...,\lambda _{n})\right.  \nonumber \\
&&\left. -B_{2}(\lambda _{1})\sum_{j=2}^{n}\frac{X_{1}(\lambda _{j}|z)}{%
y(\lambda _{1}-\lambda _{j})}\prod_{k=2,k\neq j}^{n}{\cal Z}(\lambda
_{k}-\lambda _{j})\Psi _{n-2}(\lambda _{2},...,\stackrel{\wedge }{\lambda }%
_{j},...,\lambda _{n})\right.  \label{inh.5}
\end{eqnarray}
with the initial condition $\Psi _{0}=\left| 0\right\rangle ,\quad \Psi
_{1}(\lambda _{1})=B_{1}(\lambda _{1})\left| 0\right\rangle $. \ $\stackrel{%
\wedge }{\lambda }_{j}$ denotes that the rapidity $\lambda _{j}$ is absent: $%
\Psi (\lambda _{1},...,\stackrel{\wedge }{\lambda }_{j},\cdots ,\lambda
_{n})=\Psi (\lambda _{1},...,\lambda _{j-1},\lambda _{j+1},\cdots ,\lambda
_{n})$.

Let us now describe each term which appear in the right hand side of the 
{\small OSBAE} for the $osp(1|2)$ model ( for more details the reader can
see \cite{LI}): In the first term the Bethe vectors (\ref{inh.5}) are
multiplied by $c$-numbers $\Lambda =\Lambda (\lambda ,\lambda
_{1},...,\lambda _{n}|z)$ given by 
\begin{equation}
\Lambda =X_{1}(\lambda |z)\prod_{k=1}^{n}z(\lambda _{k}-\lambda
)-(-)^{n}X_{2}(\lambda |z)\prod_{k=1}^{n}\frac{z(\lambda -\lambda _{k})}{%
\omega (\lambda -\lambda _{k})}+X_{3}(\lambda |z)\prod_{k=1}^{n}\frac{%
x_{2}(\lambda -\lambda _{k})}{x_{3}(\lambda -\lambda _{k})}.  \label{inh.6}
\end{equation}
The second term is a sum of new vectors defined as 
\begin{equation}
\Psi _{(n)}^{j}=\left( \frac{x_{5}(\lambda _{j}-\lambda )}{x_{2}(\lambda
_{j}-\lambda )}B_{1}(\lambda |z)+\frac{1}{y(\lambda -\lambda _{j})}%
B_{3}(\lambda |z)\right) \Psi _{n-1}(\stackrel{\wedge }{\lambda }_{j})
\label{inh.7}
\end{equation}
and ${\cal F}_{j}^{(n)}$ are scalar functions given by 
\begin{equation}
{\cal F}_{j}^{(n)}=X_{1}(\lambda _{j}|z)\prod_{k\neq j}^{n}{\cal Z}(\lambda
_{k}-\lambda _{j})+(-)^{n}X_{2}(\lambda _{j}|z)\prod_{k\neq j}^{n}{\cal Z}%
(\lambda _{j}-\lambda _{k}).  \label{inh.8}
\end{equation}
Finally, the last term is a coupled sum of a third type of vector-valued
functions 
\begin{equation}
\Psi _{(n)}^{jl}=B_{2}(\lambda |z)\Psi _{n-2}(\stackrel{\wedge }{\lambda }%
_{l},\stackrel{\wedge }{\lambda }_{j})  \label{inh.9}
\end{equation}
with intricate coefficients 
\begin{eqnarray}
{\cal F}_{lj}^{(n)} &=&G_{jl}X_{1}(\lambda _{l}|z)X_{1}(\lambda
_{j}|z)\prod_{k=1,k\neq j,l}^{n}{\cal Z}(\lambda _{k}-\lambda _{l}){\cal Z}%
(\lambda _{k}-\lambda _{j})  \nonumber \\
&&-(-)^{n}Y_{jl}X_{1}(\lambda _{l}|z)X_{2}(\lambda _{j}|z)\prod_{k=1,k\neq
j,l}^{n}{\cal Z}(\lambda _{k}-\lambda _{l}){\cal Z}(\lambda _{j}-\lambda
_{k})  \nonumber \\
&&-(-)^{n}F_{jl}X_{1}(\lambda _{j}|z)X_{2}(\lambda _{l}|z)\prod_{k=1,k\neq
j,l}^{n}{\cal Z}(\lambda _{l}-\lambda _{k}){\cal Z}(\lambda _{k}-\lambda
_{j})  \nonumber \\
&&+H_{jl}X_{2}(\lambda _{l}|z)X_{2}(\lambda _{j}|z)\prod_{k=1,k\neq j,l}^{n}%
{\cal Z}(\lambda _{j}-\lambda _{k}){\cal Z}(\lambda _{l}-\lambda _{k})
\label{inh.10}
\end{eqnarray}
where $G_{jl}$ , $Y_{jl}$ , $F_{jl}$ and $H_{jl}$ are additional ratio
functions defined by 
\begin{eqnarray}
G_{jl} &=&\frac{x_{7}(\lambda _{l}-\lambda )}{x_{3}(\lambda _{l}-\lambda )}%
\frac{1}{y(\lambda _{l}-\lambda _{j})}+\frac{z(\lambda _{l}-\lambda )}{%
\omega (\lambda _{l}-\lambda )}\frac{x_{5}(\lambda _{j}-\lambda )}{%
x_{2}(\lambda _{j}-\lambda )}\frac{1}{y(\lambda -\lambda _{l})}  \nonumber \\
H_{jl} &=&\frac{y_{7}(\lambda -\lambda _{l})}{x_{3}(\lambda -\lambda _{l})}%
\frac{1}{y(\lambda _{l}-\lambda _{j})}-\frac{y_{5}(\lambda -\lambda _{l})}{%
x_{3}(\lambda -\lambda _{l})}\frac{1}{y(\lambda -\lambda _{j})}  \nonumber \\
Y_{jl} &=&\frac{1}{y(\lambda -\lambda _{l})}\left\{ z(\lambda -\lambda _{l})%
\frac{y_{5}(\lambda -\lambda _{j})}{x_{2}(\lambda -\lambda _{j})}-\frac{%
y_{5}(\lambda -\lambda _{l})}{x_{2}(\lambda -\lambda _{l})}\frac{%
y_{5}(\lambda _{l}-\lambda _{j})}{x_{2}(\lambda _{l}-\lambda _{j})}\right\} 
\nonumber \\
F_{jl} &=&\frac{y_{5}(\lambda -\lambda _{l})}{x_{2}(\lambda -\lambda _{l})}%
\left\{ \frac{y_{5}(\lambda _{l}-\lambda _{j})}{x_{2}(\lambda _{l}-\lambda
_{j})}\frac{1}{y(\lambda -\lambda _{l})}+\frac{z(\lambda -\lambda _{l})}{%
\omega (\lambda -\lambda _{l})}\frac{1}{y(\lambda -\lambda _{j})}\right. 
\nonumber \\
&&\left. -\frac{y_{5}(\lambda -\lambda _{l})}{x_{2}(\lambda -\lambda _{l})}%
\frac{1}{y(\lambda _{l}-\lambda _{j})}\right\}  \label{inh.11}
\end{eqnarray}
In the usual Bethe ansatz method, the next step consist in impose the
vanishing of the so-called unwanted terms of (\ref{inh.4}) in order to get
an eigenvalue problem for the transfer matrix:

We impose ${\cal F}_{j}^{(n)}=0$ , which also implies ${\cal F}_{lj}^{(n)}$
vanishing . Hence, $\Psi _{n}(\lambda _{1},...,\lambda _{n})$ is an
eigenstate of $\tau (\lambda |z)$ with eigenvalue $\Lambda $ , provided the
rapidities $\lambda _{j}$ are solutions of the inhomogeneous Bethe ansatz
equations 
\begin{eqnarray}
\prod_{a=1}^{N}\left( \frac{\lambda _{j}-z_{a}+\eta }{\lambda _{j}-z_{a}}%
\right) &=&(-)^{n+1}\prod_{k\neq j}^{n}\frac{(\lambda _{j}-\lambda _{k}+\eta
)(\lambda _{k}-\lambda _{j}+\eta /2)}{(\lambda _{j}-\lambda _{k}-\eta
)(\lambda _{k}-\lambda _{j}-\eta /2)}  \nonumber \\
j &=&1,2,...,n  \label{inh.12}
\end{eqnarray}
These equations were derived for the first time in \cite{MR}.

\section{Structure of the ${\bf osp(1|2)}$ Gaudin Model}

In this section we will consider the theory of the Gaudin model. To do this
we need to calculate the semiclassical limit of the results presented in the
previous section.

In order to expand the matrix elements of $T(\lambda |z)$, up to an
appropriate order in $\eta $, we will start by expanding the Lax operator
entries defined in (\ref{gra.7}): 
\begin{eqnarray}
L_{11}^{(n)}(\lambda |z) &=&1+\eta \frac{2H_{n}}{\lambda -z_{n}}+\frac{3}{2}%
\eta ^{2}\frac{2H_{n}^{2}+H_{n}}{(\lambda -z_{n})^{2}}+o(\eta ^{3}), 
\nonumber \\
L_{22}^{(n)}(\lambda |z) &=&1+\eta \ 0+\frac{3}{2}\eta ^{2}\frac{%
1_{n}-4H_{n}^{2}}{(\lambda -z_{n})^{2}}+o(\eta ^{3}),  \nonumber \\
L_{33}^{(n)}(\lambda |z) &=&1-\eta \frac{2H_{n}}{\lambda -z_{n}}+\frac{3}{2}%
\eta ^{2}\frac{2H_{n}^{2}-H_{n}}{(\lambda -z_{n})^{2}}+o(\eta ^{3}),
\label{gau.1}
\end{eqnarray}
and for off-diagonal elements, we have 
\begin{eqnarray}
L_{12}^{(n)}(\lambda |z) &=&-\eta \frac{2V_{n}^{-}}{\lambda -z_{n}}+o(\eta
^{2}),\qquad \ L_{21}^{(n)}(\lambda |z)=\eta \frac{2V_{n}^{+}}{\lambda -z_{n}%
}+o(\eta ^{2}),  \nonumber \\
L_{13}^{(n)}(\lambda |z) &=&\eta \frac{2X_{n}^{-}}{\lambda -z_{n}}+o(\eta
^{2}),\qquad \quad L_{31}^{(n)}(\lambda |z)=\eta \frac{2X_{n}^{+}}{\lambda
-z_{n}}+o(\eta ^{2}),  \nonumber \\
L_{23}^{(n)}(\lambda |z) &=&\eta \frac{2V_{n}^{-}}{\lambda -z_{n}}+o(\eta
^{2}),\qquad \quad L_{32}^{(n)}(\lambda |z)=\eta \frac{2V_{n}^{+}}{\lambda
-z_{n}}+o(\eta ^{2}).  \label{gau.2}
\end{eqnarray}
Here $H$, $X^{\pm }$ and $V^{\pm }$ are matrices $3$ by $3$ given by (\ref
{str.4}).

Using (\ref{gau.1}) and (\ref{gau.2}) is now easy to derive the
semiclassical expansion of the\ monodromy matrix entries: 
\begin{eqnarray}
A_{1}(\lambda |z) &=&\Gamma (\lambda |z)\left\{ 1+\eta \sum_{a=1}^{N}\frac{%
{\cal H}_{a}}{\lambda -z_{a}}\right.  \nonumber \\
&&+\!\left. \eta ^{2}\sum_{a<b}\left( \!\frac{{\cal H}_{a}\stackrel{s}{%
\otimes }{\cal H}_{b}+{\cal X}_{a}^{-}\stackrel{s}{\otimes }{\cal X}_{b}^{+}-%
{\cal V}_{a}^{-}\stackrel{s}{\otimes }{\cal V}_{b}^{+}}{(\lambda
-z_{a})(\lambda -z_{b})}\!+\!\frac{3}{4}\sum_{a=1}^{N}\frac{[{\cal H}^{2}-%
{\cal H}]_{a}}{(\lambda -z_{a})^{2}}\right) \!+o(\eta ^{3})\right\} 
\nonumber \\
A_{2}(\lambda |z) &=&\Gamma (\lambda |z)\left\{ 1+\eta \ 0\right.  \nonumber
\\
&&+\left. \eta ^{2}\sum_{a<b}\left( \frac{{\cal V}_{a}^{-}\stackrel{s}{%
\otimes }{\cal V}_{b}^{+}-{\cal V}_{a}^{+}\stackrel{s}{\otimes }{\cal V}%
_{b}^{-}}{(\lambda -z_{a})(\lambda -z_{b})}-\frac{3}{2}\sum_{a=1}^{N}\frac{%
[1-{\cal H}^{2}]_{a}}{(\lambda -z_{a})^{2}}\right) +o(\eta ^{3})\right\} 
\nonumber \\
A_{3}(\lambda |z) &=&\Gamma (\lambda |z)\left\{ 1-\eta \sum_{a=1}^{N}\frac{%
{\cal H}_{a}}{\lambda -z_{a}}\right.  \nonumber \\
&&+\!\left. \eta ^{2}\sum_{a<b}\left( \!\frac{{\cal H}_{a}\stackrel{s}{%
\otimes }{\cal H}_{b}+{\cal X}_{a}^{+}\stackrel{s}{\otimes }{\cal X}_{b}^{-}+%
{\cal V}_{a}^{+}\stackrel{s}{\otimes }{\cal V}_{b}^{-}}{(\lambda
-z_{a})(\lambda -z_{b})}\!+\!\frac{3}{4}\sum_{a=1}^{N}\frac{[{\cal H}^{2}+%
{\cal H}]_{a}}{(\lambda -z_{a})^{2}}\right) \!+o(\eta ^{3})\right\} 
\nonumber \\
B_{1}(\lambda |z) &=&-B_{3}(\lambda |z)=\Gamma (\lambda |z)\left\{ -\eta
\sum_{a=1}^{N}\frac{{\cal V}_{a}^{-}}{\lambda -z_{a}}+o(\eta ^{2})\right\} 
\nonumber \\
C_{1}(\lambda |z) &=&C_{3}(\lambda |z)=\Gamma (\lambda |z)\left\{ \eta
\sum_{a=1}^{N}\frac{{\cal V}_{a}^{+}}{\lambda -z_{a}}+o(\eta ^{2})\right\} 
\nonumber \\
B_{2}(\lambda |z) &=&\Gamma (\lambda |z)\left\{ \eta \sum_{a=1}^{N}\frac{%
{\cal X}_{a}^{-}}{\lambda -z_{a}}+o(\eta ^{2})\right\}  \nonumber \\
C_{2}(\lambda |z) &=&\Gamma (\lambda |z)\left\{ \eta \sum_{a=1}^{N}\frac{%
{\cal X}_{a}^{+}}{\lambda -z_{a}}+o(\eta ^{2})\right\}  \label{gau.3}
\end{eqnarray}
where ${\cal H}=2H,\ {\cal X}^{\pm }=2X^{\pm },\ {\cal V}^{\pm }=2V^{\pm }\ $%
and $\Gamma (\lambda |z)=\prod_{a=1}^{N}(\lambda -z_{a})$ is a common factor
which can be absorbed after a convenient normalization. From the definition (%
\ref{gra.10}) it follows that the semiclassical expansion of the normalized
transfer matrix $t(\lambda |z)=\tau (\lambda |z)/\Gamma (\lambda |z)$ has
the form 
\begin{equation}
t(\lambda |z)=1+2\eta ^{2}\left\{ \sum_{a=1}^{N}\frac{G_{a}}{\lambda -z_{a}}+%
\frac{3}{4}\sum_{a=1}^{N}\frac{[1]_{a}}{(\lambda -z_{a})^{2}}\right\}
+o(\eta ^{3})  \label{gau.5}
\end{equation}
where 
\begin{equation}
G_{a}=\sum_{b\neq a}^{N}\frac{{\cal H}_{a}\stackrel{s}{\otimes }{\cal H}_{b}+%
\frac{1}{2}({\cal X}_{a}^{+}\stackrel{s}{\otimes }{\cal X}_{b}^{-}+{\cal X}%
_{a}^{-}\stackrel{s}{\otimes }{\cal X}_{b}^{+})+{\cal V}_{a}^{+}\stackrel{s}{%
\otimes }{\cal V}_{b}^{-}-{\cal V}_{a}^{-}\stackrel{s}{\otimes }{\cal V}%
_{b}^{+}}{z_{a}-z_{b}}  \label{gau.6}
\end{equation}
Here we have used the symmetry $G_{kj}={\cal P}_{jk}G_{jk}{\cal P}%
_{jk}=G_{jk}$ and the identity 
\begin{equation}
\frac{1}{(\lambda -z_{a})(\lambda -z_{b})}=\left( \frac{1}{\lambda -z_{a}}-%
\frac{1}{\lambda -z_{b}}\right) \frac{1}{z_{a}-z_{b}}  \label{gau.7}
\end{equation}

The Gaudin model is defined by the residue of $t(\lambda |z)$ (\ref{gau.5})
at the point $\lambda =z_{a}$. This results in $N$ non-local magnets whose
Hamiltonians $G_{a}$ , $a=1,2,...,N$ are given by (\ref{gau.6}) and
satisfied the property $\sum_{a=1}^{N}G_{a}=0$.

\subsection{${\bf osp(1|2)}$ Gaudin Algebra}

Let us now consider the semiclassical limit of the fundamental commutation
relation (\ref{gra.5}). Let us first write the $R$-matrix (\ref{gra.3}) in
the form \cite{Ku2}: 
\begin{equation}
R(\lambda ,\eta )=\eta {\cal I}+\lambda {\cal P}+\frac{\lambda \eta }{%
\lambda +\frac{3}{2}\eta }{\cal U}  \label{alg.1}
\end{equation}
where ${\cal I}$ is the identity operator, ${\cal P}$ is the graded
permutation operator (\ref{gra.2}) and ${\cal U}$ is the rank-one projector $%
{\cal U}^{2}={\cal U}$. Using the normalization of the previous section we
can write the semiclassical expansions of $T$ and $R$ in the following form 
\begin{eqnarray}
T(\lambda |z) &=&1+\eta L(\lambda |z)+o(\eta ^{2})  \nonumber \\
R(\lambda -\mu ) &=&{\cal P}\left[ 1+\eta r(\lambda -\mu )+o(\eta ^{2})%
\right]  \label{alg.2}
\end{eqnarray}
From (\ref{gau.3}) we can see that the \ ''classical $L$-operator '' has the
form 
\begin{equation}
L(\lambda |z)=\left( 
\begin{array}{ccc}
{\cal H}(\lambda |z) & -{\cal V}^{-}(\lambda |z) & {\cal X}^{-}(\lambda |z)
\\ 
{\cal V}^{+}(\lambda |z) & 0 & {\cal V}^{-}(\lambda |z) \\ 
{\cal X}^{+}(\lambda |z) & {\cal V}^{+}(\lambda |z) & -{\cal H}(\lambda |z)
\end{array}
\right)  \label{alg.3}
\end{equation}
where 
\begin{eqnarray}
{\cal H}(\lambda |z) &=&\sum_{a=1}^{N}\frac{{\cal H}_{a}}{\lambda -z_{a}}%
,\quad {\cal V}^{-}(\lambda |z)=\sum_{a=1}^{N}\frac{{\cal V}_{a}^{-}}{%
\lambda -z_{a}},\quad {\cal X}^{-}(\lambda |z)=\sum_{a=1}^{N}\frac{{\cal X}%
_{a}^{-}}{\lambda -z_{a}}  \nonumber \\
{\cal V}^{+}(\lambda |z) &=&\sum_{a=1}^{N}\frac{{\cal V}_{a}^{+}}{\lambda
-z_{a}},\quad {\cal X}^{+}(\lambda |z)=\sum_{a=1}^{N}\frac{{\cal X}_{a}^{+}}{%
\lambda -z_{a}}  \label{alg.4}
\end{eqnarray}
and the classical $r$-matrix is obtained by the semiclassical expansion of (%
\ref{alg.1}). It has the form 
\begin{equation}
r(\lambda -\mu )=\frac{1}{\lambda -\mu }\left( {\cal P}-{\cal U}\right)
\label{alg.5}
\end{equation}

Substituting (\ref{alg.3}) and (\ref{alg.4}) into the fundamental relation (%
\ref{gra.5}) we will get, from the first non trivial identity , the
following equation: 
\begin{eqnarray}
&&{\cal P}L(\lambda |z)\stackrel{s}{\otimes }L(\mu |z)+{\cal P}r(\lambda
-\mu )\left[ L(\lambda |z)\stackrel{s}{\otimes }1+1\stackrel{s}{\otimes }%
L(\mu |z)\right]  \nonumber \\
&=&L(\mu |z)\stackrel{s}{\otimes }L(\lambda |z){\cal P}+\left[ L(\lambda |z)%
\stackrel{s}{\otimes }1+1\stackrel{s}{\otimes }L(\mu |z)\right] {\cal P}%
r(\lambda -\mu )  \label{alg.6}
\end{eqnarray}
whose consistence is guaranteed by the graded classical {\small YB} equation.

From (\ref{alg.6}) we can derive commutations relations between the matrix
elements of $L(\lambda |z)$. They are the defining relations of the $%
osp(1|2) $ Gaudin algebra presented in \cite{BM}: 
\begin{eqnarray}
\lbrack {\cal H}(\lambda |z),{\cal H}(\mu |z)] &=&[{\cal V}^{-}(\lambda |z),%
{\cal X}^{-}(\mu |z)]=[{\cal X}^{-}(\lambda |z),{\cal X}^{-}(\mu |z)]=0 
\nonumber \\
\quad \lbrack {\cal V}^{+}(\lambda |z),{\cal X}^{+}(\mu |z)] &=&[{\cal X}%
^{+}(\lambda |z),{\cal X}^{+}(\mu |z)]=0  \nonumber \\
\lbrack {\cal H}(\lambda |z),{\cal V}^{-}(\mu |z)] &=&-2[{\cal X}%
^{-}(\lambda |z),{\cal V}^{+}(\mu |z)]=\frac{1}{\lambda -\mu }[{\cal V}%
^{-}(\lambda |z)-{\cal V}^{-}(\mu |z)]  \nonumber \\
\lbrack {\cal H}(\lambda |z),{\cal X}^{-}(\mu |z)] &=&2\{{\cal V}%
^{-}(\lambda |z),{\cal V}^{-}(\mu |z)\}=\frac{2}{\lambda -\mu }[{\cal X}%
^{-}(\lambda |z)-{\cal X}^{-}(\mu |z)]  \nonumber \\
\lbrack {\cal H}(\lambda |z),{\cal V}^{+}(\mu |z)] &=&-2[{\cal V}%
^{-}(\lambda |z),{\cal X}^{+}(\mu |z)]=-\frac{1}{\lambda -\mu }[{\cal V}%
^{+}(\lambda |z)-{\cal V}^{+}(\mu |z)]  \nonumber \\
\lbrack {\cal H}(\lambda |z),{\cal X}^{+}(\mu |z)] &=&2\{{\cal V}%
^{+}(\lambda |z),{\cal V}^{+}(\mu |z)\}=-\frac{2}{\lambda -\mu }[{\cal X}%
^{+}(\lambda |z)-{\cal X}^{+}(\mu |z)]  \nonumber \\
\lbrack {\cal X}^{-}(\lambda |z),{\cal X}^{+}(\mu |z)] &=&4\{{\cal V}%
^{-}(\lambda |z),{\cal V}^{+}(\mu |z)\}=\frac{4}{\lambda -\mu }[{\cal H}%
(\lambda |z)-{\cal H}(\mu |z)]  \label{alg.7}
\end{eqnarray}
It follows from (\ref{alg.6}) \ that the Gaudin algebra is the semiclassical
limit of the commutation relations of the matrix elements of the monodromy
matrix used in the usual Bethe ansatz method.

A direct consequence of these relations is the commutativity of $t(\lambda
|z)$ (\ref{gau.5}) 
\begin{equation}
\lbrack t(\lambda |z),t(\mu |z)]=0,\qquad \forall \lambda ,\mu  \label{alg.8}
\end{equation}
and as consequence it follows the commutativity of the Gaudin Hamiltonians $%
G_{a}$ $,a=1,...,N$.

\subsection{Gaudin Off-shell Bethe Ansatz Equations}

In order to get semiclassical limit of the {\small OSBAE} (\ref{inh.4}) we
first consider the semiclassical expansions of the Bethe vectors defined in (%
\ref{inh.5}), (\ref{inh.7}) and (\ref{inh.9}): 
\begin{eqnarray}
\Psi _{n}(\lambda _{1},...,\lambda _{n}) &=&(-\eta )^{n}\Phi _{n}(\lambda
_{1},...,\lambda _{n}|z)+o(\eta ^{n+1})  \nonumber \\
\Psi _{(n)}^{j} &=&2(-\eta )^{n+1}\frac{{\cal V}^{-}(\lambda |z)}{\lambda
-\lambda _{j}}\Phi _{n-1}(\lambda _{1},...,\stackrel{\wedge }{\lambda }%
_{j},...,\lambda _{n}|z)+o(\eta ^{n+2})  \nonumber \\
\Psi _{(n)}^{jl} &=&-(-\eta )^{n-1}{\cal X}^{-}(\lambda |z)\Phi
_{n-2}(\lambda _{1},...,\stackrel{\wedge }{\lambda }_{l},\stackrel{\wedge }{%
\lambda }_{j},...,\lambda _{n}|z)+o(\eta ^{n})  \label{off.1}
\end{eqnarray}
where 
\begin{eqnarray}
\Phi _{n}(\lambda _{1},...,\lambda _{n}|z) &=&{\cal V}^{-}(\lambda
_{1}|z)\Phi _{n-1}(\lambda _{2},...,\lambda _{n}|z)  \nonumber \\
&&-{\cal X}^{-}(\lambda _{1}|z)\sum_{j=2}^{n}\frac{(-)^{j}}{\lambda
_{1}-\lambda _{j}}\Phi _{n-2}(\lambda _{2},...,\stackrel{\wedge }{\lambda }%
_{j},...,\lambda _{n}|z)  \label{off.2}
\end{eqnarray}
with $\Phi _{0}=\left| 0\right\rangle $ and $\Phi _{1}(\lambda _{1}|z)={\cal %
V}^{-}(\lambda _{1}|z)\left| 0\right\rangle $.

Here we would like make a few comments on the structure of these
vector-valued functions. In (\ref{off.2}) they are written in a normal
ordered form. Since we are working with fermionic degree of freedom, the
function $\Phi _{n}(\lambda _{1},...,\lambda _{n}|z)$ is totally
antisymmetric. 
\begin{equation}
\Phi _{n}(\lambda _{1},...,\lambda _{i-1},\lambda _{i+1},\lambda
_{i},...,\lambda _{n}|z)=-\Phi _{n}(\lambda _{1},...,\lambda _{i-1},\lambda
_{i},\lambda _{i+1},...,\lambda _{n}|z)  \label{off.3}
\end{equation}
To see this one can use the Gaudin algebra (\ref{alg.7}). For instance, in
its antisymmetric form the Bethe vector $\Phi _{2}$ reads as 
\begin{eqnarray}
\Phi _{2}(\lambda _{1},\lambda _{2}|z) &=&\frac{1}{2}[{\cal V}^{-}(\lambda
_{1}|z){\cal V}^{-}(\lambda _{2}|z)-{\cal V}^{-}(\lambda _{2}|z){\cal V}%
^{-}(\lambda _{1}|z)]\left| 0\right\rangle  \nonumber \\
&&-\frac{1}{2}[\frac{{\cal X}^{-}(\lambda _{1}|z)+{\cal X}^{-}(\lambda
_{2}|z)}{\lambda _{1}-\lambda _{2}}]\left| 0\right\rangle  \label{off.4}
\end{eqnarray}

Next we consider the semiclassical expansions of the $c$-number functions (%
\ref{inh.6}), (\ref{inh.8}) and (\ref{inh.10}) presents in the {\small OSBAE}
(\ref{inh.4}) \ 
\begin{eqnarray}
\Lambda &=&1+2\eta ^{2}\Lambda ^{(2)}+o(\eta ^{3})  \nonumber \\
{\cal F}_{f}^{(n)} &=&\eta (-)^{j}f_{j}^{(n)}+o(\eta ^{2}),\quad {\cal F}%
_{lj}^{(n)}=2\eta ^{3}(-)^{j+l}F_{lj}^{(n)}+o(\eta ^{4})  \label{off.5}
\end{eqnarray}
where 
\begin{eqnarray}
\Lambda ^{(2)} &=&\sum_{a=1}^{N}\frac{1}{(\lambda -z_{a})}\left( \sum_{b\neq
a}^{N}\frac{1}{z_{a}-z_{b}}\right) +\sum_{k=1}^{n}\frac{1}{(\lambda -\lambda
_{k})}\left( \sum_{j\neq k}^{n}\frac{1}{\lambda _{k}-\lambda _{j}}\right) 
\nonumber \\
&&+\sum_{a=1}^{N}\sum_{k=1}^{n}\frac{1}{\lambda -z_{a}}\frac{1}{\lambda
_{k}-\lambda }+\frac{3}{4}\sum_{a=1}^{N}\frac{1}{(\lambda -z_{a})^{2}}
\label{off.6}
\end{eqnarray}
and 
\begin{eqnarray}
f_{j}^{(n)} &=&\sum_{k\neq j}^{n}\frac{1}{\lambda _{j}-\lambda _{k}}%
-\sum_{a=1}^{N}\frac{1}{\lambda _{j}-z_{a}}  \nonumber \\
F_{lj}^{(n)} &=&\frac{f_{j}^{(n)}}{\lambda -\lambda _{j}}\frac{1}{\lambda
_{j}-\lambda _{l}}-\frac{f_{l}^{(n)}}{\lambda -\lambda _{l}}\frac{1}{\lambda
_{l}-\lambda _{j}}  \label{off.7}
\end{eqnarray}
Substituting these expressions into the (\ref{inh.4}) and comparing the
coefficients of the terms $2(-\eta )^{n+2}$ we get the first non-trivial
consequence for the semiclassical limit of the \ {\small OSBAE}: 
\begin{equation}
\sum_{a=1}^{N}\frac{G_{a}}{\lambda -z_{a}}\Phi _{n}(\lambda _{1},...,\lambda
_{n})=\Lambda ^{(2)}\Phi _{n}(\lambda _{1},...,\lambda _{n})+\sum_{j=1}^{n}%
\frac{(-)^{l}f_{l}^{(n)}\Phi _{n}^{j}}{\lambda -\lambda _{j}}  \label{off.8}
\end{equation}
where 
\begin{eqnarray}
\Phi _{n}^{j} &=&{\cal V}^{-}(\lambda |z)\Phi _{n-1}(\lambda _{1},...,%
\stackrel{\wedge }{\lambda }_{j},...,\lambda _{n}|z)  \nonumber \\
&&-{\cal X}^{-}(\lambda |z)\sum_{l\neq j}^{n}\frac{(-)^{\stackrel{\sim }{l}}%
}{\lambda _{j}-\lambda _{l}}\Phi _{n-2}(\lambda _{1},...,\stackrel{\wedge }{%
\lambda }_{l},...,\stackrel{\wedge }{\lambda }_{j},...,\lambda _{n}|z)
\label{off.9}
\end{eqnarray}
with $\stackrel{\sim }{l}=l+1$ \ for {\rm \ }$l<j$ or $\stackrel{\sim }{l}=l$
$\ $for {\rm \ }$l>j$.

Here we observe that the {\small OSBAE} (\ref{off.8}) is similar to the 
{\small OSBAE} presented by Babujian and Flume for simple Lie algebras. This
result could be expected since the superalgebra $osp(1|2)$ has many features
which make \ it very close to the Lie algebra \cite{FSS}.

Now, we take the residue of (\ref{off.8}) at the point $\lambda =z_{a}$ to
get the Gaudin off-shell equations: 
\begin{eqnarray}
G_{a}\Phi _{n}(\lambda _{1},...,\lambda _{n}|z) &=&g_{a}\Phi _{n}(\lambda
_{1},...,\lambda _{n}|z)+\sum_{l=1}^{n}\frac{(-)^{l}f_{l}^{(n)}\phi _{n}^{l}%
}{z_{a}-\lambda _{l}}  \nonumber \\
a &=&1,2,...,N  \label{off.11}
\end{eqnarray}
where $f_{l}^{(n)}$ is given by (\ref{off.7}) , $g_{a}$ is the residue of $%
\Lambda ^{(2)}$ (\ref{off.6}) at $\lambda =z_{a}$ 
\begin{equation}
g_{a}=\sum_{b\neq a}^{N}\frac{1}{z_{a}-z_{b}}-\sum_{l=1}^{n}\frac{1}{%
z_{a}-\lambda _{l}}  \label{off.12}
\end{equation}
and $\phi _{n}^{l}$ is the residue of $\Phi _{n}^{l}$ (\ref{off.9}) at $%
\lambda =z_{a}$ 
\begin{equation}
\phi _{n}^{j}={\cal V}_{j}^{-}\Phi _{n-1}(\stackrel{\wedge }{\lambda }%
_{j}|z)-{\cal X}_{j}^{-}\sum_{l\neq j}^{n}\frac{(-)^{\stackrel{\sim }{l}}}{%
\lambda _{j}-\lambda _{l}}\Phi _{n-2}(\stackrel{\wedge }{\lambda }_{l},%
\stackrel{\wedge }{\lambda }_{j}|z)  \label{off.13}
\end{equation}
In this way we arrive to one of the main problem \ solved by the theory of
the Gaudin model, {\it i.e.}, the determination of the eigenvalues and
eigenvectors of the commuting Hamiltonians $G_{a}$ (\ref{gau.6}): $g_{a}$ is
the eigenvalue of $G_{a}$ with eigenfunction $\Phi _{n}$ provided $\lambda
_{l}$ are solutions of the Bethe ansatz equations 
\begin{equation}
f_{l}^{(n)}=0\Rightarrow \sum_{k\neq l}^{n}\frac{1}{\lambda _{l}-\lambda _{k}%
}=\sum_{a=1}^{N}\frac{1}{\lambda _{l}-z_{a}},\quad l=1,2,...,n.
\label{off.14}
\end{equation}
These Bethe ansatz equations are the semiclassical limit of the
inhomogeneous Bethe ansatz equations (\ref{inh.12}).

\section{Knizhnik-Zamolodchickov Equation}

Since the spectra of the $osp(1|2)$ Gaudin Hamiltonians are now known, the
next step is to calculate correlation functions.

Before we present the result of this section, let us observe a very
important consequence due to semiclassical approach. While the solutions of
the quantum {\small YB} equation (\ref{int.1}) depend essentially on the
representation, the corresponding classical solutions $r_{ij}(\lambda )\in 
{\rm g}_{i}\otimes {\rm g}_{j}$ , ${\rm g}$ being a Lie algebra, can be
written in an invariant form, i.e. independent of the representation.
Therefore, one may extend the previous semiclassical results beyond the
fundamental representation. To do this we first define a $r$-matrix and the
corresponding $L$-operator for the next representation ($s=1$) of the $%
osp(1|2)$ algebra presented in (\ref{str.3}). The classical $r$-matrix is
constructed out of the quadratic Casimir in a standard way \cite{Ku2}, 
\begin{equation}
r(\lambda -\mu )=\frac{1}{\lambda -\mu }[{\cal H}\stackrel{s}{\otimes }{\cal %
H}+\frac{1}{2}({\cal X}^{+}\stackrel{s}{\otimes }{\cal X}^{-}+{\cal X}^{-}%
\stackrel{s}{\otimes }{\cal X}^{+})+{\cal V}^{+}\stackrel{s}{\otimes }{\cal V%
}^{-}-{\cal V}^{-}\stackrel{s}{\otimes }{\cal V}^{+}]  \label{kz.2}
\end{equation}
Using the basis (\ref{str.3}) with the grading {\small BFBFB} one can verify
that this classical $r$-matrix satisfies the graded version of the classical 
{\small YB} equation (\ref{int.2}). To write (\ref{alg.5}) in the form (\ref
{kz.2}), we recall the symmetries of the {\small YB} solution (\ref{gra.4})
by considering the case $\epsilon _{1}=1$ and $\epsilon _{2}=-1$.

The classical $L$-operator is obtained from the semiclassical limit of the
canonical identification of the Lax operator ${\cal L}$ with ${\cal R}$%
-matrix \cite{FT,FD}. In the $s=1$ representation it has the form 
\begin{equation}
L(\lambda |z)=\left( 
\begin{array}{ccccc}
2{\cal H}(\lambda |z) & \sqrt{2}{\cal V}^{-}(\lambda |z) & \sqrt{2}{\cal X}%
^{-}(\lambda |z) & 0 & 0 \\ 
-\sqrt{2}{\cal V}^{+}(\lambda |z) & {\cal H}(\lambda |z) & {\cal V}%
^{-}(\lambda |z) & {\cal X}^{-}(\lambda |z) & 0 \\ 
\sqrt{2}{\cal X}^{+}(\lambda |z) & {\cal V}^{+}(\lambda |z) & 0 & {\cal V}%
^{-}(\lambda |z) & \sqrt{2}{\cal X}^{-}(\lambda |z) \\ 
0 & {\cal X}^{+}(\lambda |z) & -{\cal V}^{+}(\lambda |z) & -{\cal H}(\lambda
|z) & \sqrt{2}{\cal V}^{-}(\lambda |z) \\ 
0 & 0 & \sqrt{2}{\cal X}^{+}(\lambda |z) & \sqrt{2}{\cal V}^{+}(\lambda |z)
& -2{\cal H}(\lambda |z)
\end{array}
\right)  \label{kz.1}
\end{equation}

Substituting (\ref{kz.1}) and (\ref{kz.2}) into the fundamental relation (%
\ref{alg.6}) we will get the same defining relations of the $osp(1|2)$
Gaudin algebra (\ref{alg.7}).

The fundamental property of the operators (\ref{alg.4}) is that they form
the highest weight module of the infinite-dimensional Lie superalgebra. As
in the $sl(2)$ case presented by Sklyanin in \cite{SKL} , it is
characterized by the vacuum $\left| 0\right\rangle $ , its dual $%
\left\langle 0\right| $%
\begin{eqnarray}
{\cal H}(\lambda |z)\left| 0\right\rangle &=&h(\lambda |z)\left|
0\right\rangle ,\qquad {\cal V}^{+}(\lambda |z)\left| 0\right\rangle
=0,\qquad {\cal X}^{+}(\lambda |z)\left| 0\right\rangle =0  \nonumber \\
\left\langle 0\right| {\cal H}(\lambda |z) &=&h(\lambda |z)\left\langle
0\right| ,\qquad \left\langle 0\right| {\cal V}^{-}(\lambda |z)=0,\qquad
\left\langle 0\right| {\cal X}^{-}(\lambda |z)=0  \label{kz.3}
\end{eqnarray}
and the highest weight functions $h(\lambda |z)$%
\begin{equation}
h(\lambda |z)=\sum_{a=1}^{N}\frac{2s_{a}}{\lambda -z_{a}},\qquad s_{a}=\frac{%
1}{2},1,\frac{3}{2},...  \label{kz.4}
\end{equation}
In this way we are going beyond the fundamental representation $s_{a}=\frac{1%
}{2}$.

The semiclassical expansion of $t(\lambda |z)$ in the $s$-representation is
easily obtained from (\ref{gau.5}) 
\begin{equation}
t(\lambda |z)=1+2\eta ^{2}\left\{ \sum_{a=1}^{N}\frac{G_{a}}{\lambda -z_{a}}%
+\sum_{a=1}^{N}\frac{s_{a}(s_{a}+1)}{(\lambda -z_{a})^{2}}\right\} +o(\eta
^{3})  \label{kz.5}
\end{equation}
Here we remark that this result is general. It comes from the structure of
the semiclassical method, where $t(\lambda |z)$ is defined as supertrace of $%
\frac{1}{2}L^{2}(\lambda |z)$ and the second sum in (\ref{kz.5}) is
identified with the Casimir of the subalgebra $sl(2)$. Consequently, the
Gaudin off-shell equations in the $s$-representation are known. They are
again given by (\ref{off.11}), where the coefficients $f_{l}^{(n)}$ and $%
g_{a}$ are 
\begin{eqnarray}
f_{l}^{(n)} &=&\sum_{k\neq l}^{n}\frac{1}{\lambda _{l}-\lambda _{k}}%
-2\sum_{a=1}^{N}\frac{s_{a}}{\lambda _{l}-z_{a}}  \nonumber \\
g_{a} &=&4\sum_{b\neq a}^{N}\frac{s_{a}s_{b}}{z_{a}-z_{b}}-2\sum_{l=1}^{n}%
\frac{s_{a}}{z_{a}-\lambda _{l}}  \label{kz.6}
\end{eqnarray}
and $\Phi _{n}(\lambda _{1},...,\lambda _{n})$ for $s$-representations again
has the form as Eq.(\ref{off.2}).

Let us now define the vector-valued function $\Psi (z_{1},...,z_{N})$
through multiple contour integrals of the Bethe vectors (\ref{off.2}) 
\begin{equation}
\Psi (z_{1},...,z_{N})=\oint \cdots \oint {\cal X}(\lambda |z)\Phi
_{n}(\lambda |z)d\lambda _{1}...d\lambda _{n}  \label{kz.7}
\end{equation}
where ${\cal X}$ $(\lambda |z)={\cal X}$ $(\lambda _{1},...,\lambda
_{n},z_{1},...,z_{N})$ is a scalar function which in this stage is still
undefined.

In analogy with the $sl(2)$ case \cite{BAF}, we assume that $\Psi
(z_{1},...,z_{N})$ is a solution of the equations 
\begin{equation}
\kappa \frac{\partial \Psi (z_{1},...,z_{N})}{\partial z_{a}}=G_{a}\Psi
(z_{1},...,z_{N}),\quad a=1,2,...,N  \label{kz.8}
\end{equation}
where $G_{a}$ are the Gaudin Hamiltonians (\ref{gau.6}) and $\kappa $ is a
constant.

By construction, these equations are the graded version of the {\small KZ}
equations \cite{Di, Fuchs}. Hence, we can interpret $\Psi (z_{1},...,z_{N})$
as a integral representation for the $N$-point correlation function in an
chiral conformal field theory with $osp(1|2)$ symmetry \cite{Nam, KIRI}.

Substituting (\ref{kz.7}) into (\ref{kz.8}) we have 
\begin{equation}
\kappa \frac{\partial \Psi (z_{1},...,z_{N})}{\partial z_{a}}=\oint \cdots
\oint [\kappa \frac{\partial {\cal X}(\lambda |z)}{\partial z_{a}}\Phi
_{n}(\lambda |z)+\kappa {\cal X}(\lambda |z)\frac{\partial \Phi _{n}(\lambda
|z)}{\partial z_{a}}]d\lambda _{1}...d\lambda _{n}  \label{kz.9}
\end{equation}
From (\ref{off.2}) and using the Gaudin algebra (\ref{alg.3}) one can derive
the following identity 
\begin{equation}
\frac{\partial \Phi _{n}}{\partial z_{a}}=\sum_{l=1}^{n}(-)^{l}\frac{%
\partial }{\partial \lambda _{l}}\left( \frac{\phi _{n}^{l}}{\lambda
_{l}-z_{a}}\right)  \label{kz.10}
\end{equation}
which allows us write (\ref{kz.9}) in the form 
\begin{eqnarray}
\kappa \frac{\partial \Psi }{\partial z_{a}} &=&\oint \cdots \oint [\kappa 
\frac{\partial {\cal X}(\lambda |z)}{\partial z_{a}}\Phi _{n}(\lambda
|z)-\sum_{l=1}^{n}(-)^{l}\kappa \frac{\partial {\cal X}(\lambda |z)}{%
\partial \lambda _{l}}\left( \frac{\phi _{n}^{l}}{\lambda _{l}-z_{a}}\right)
]d\lambda _{1}...d\lambda _{n}  \nonumber \\
&&+\kappa \oint \cdots \oint [\sum_{l=1}^{n}(-)^{l}\frac{\partial }{\partial
\lambda _{l}}\left( \frac{{\cal X}(\lambda |z)\phi _{n}^{l}}{\lambda
_{l}-z_{a}}\right) ]d\lambda _{1}...d\lambda _{n}  \label{kz.11}
\end{eqnarray}
It is evident that the last term of (\ref{kz.11}) is vanishes, because the
contours are closed. Moreover, if the scalar function ${\cal X}(\lambda |z)$
satisfies the following differential equations 
\begin{equation}
\kappa \frac{\partial {\cal X}(\lambda |z)}{\partial z_{a}}=g_{a}{\cal X}%
(\lambda |z),\qquad \kappa \frac{\partial {\cal X}(\lambda |z)}{\partial
\lambda _{j}}=f_{j}^{(n)}{\cal X}(\lambda |z),  \label{kz.12}
\end{equation}
we are recovering the off-shell Bethe ansatz equations for the Gaudin
magnets (\ref{off.11}) from the first term in (\ref{kz.11}). Taking into
account (\ref{kz.6}) we can see that the consistency condition of the system
(\ref{kz.12}) is insured by the zero curvature conditions $\partial
f_{j}^{(n)}/\partial z_{a}=\partial g_{a}/\partial \lambda _{j}$.

The solution of (\ref{kz.12}) is 
\begin{equation}
{\cal X}(\lambda |z)=\prod_{a<b}^{N}(z_{a}-z_{b})^{4s_{a}s_{b}/\kappa
}\prod_{j<k}^{n}(\lambda _{j}-\lambda _{k})^{1/\kappa
}\prod_{a,j}^{N,n}(z_{a}-\lambda _{j})^{-2s_{a}/\kappa }.  \label{kz.13}
\end{equation}
This function determines the monodromy of $\Psi (z_{1},...,z_{N})$ as
solution of the $osp(1|2)$ {\small KZ} equation (\ref{kz.8}) and these
results are in agreement with the Schechtman-Varchenko construction for
multiple contour integral as solutions of the KZ equation in an arbitrary
simple Lie algebra \cite{SV}.

\section{Conclusion}

In the first part of this paper we have considered an integrable
representation of the $osp(1|2)$ algebra to study the algebraic Bethe ansatz
for an inhomogeneous vertex model.

In the second part, through the semiclassical limit of the quantum inverse
scattering equations we got arrive to the theory of the Gaudin model. Here,
with aid of the Gaudin algebra, we extended the results presented in the
first part for highest representations and find the spectra of $N$
commutating Hamiltonians.

We believe that the procedure presented here will work for all models for
which the usual algebraic Bethe ansatz is known. Therefore, it should be
interesting extend this procedure for all algebraic versions of the Bethe
ansatz. In particular, the natural candidate will be the nested algebraic
Bethe ansatz \cite{deVega}. In \cite{BKZ}, a system of matrix difference
equations was solved by means of a nested version of a off-shell Bethe
ansatz.

Regarding our results about the $N$-point correlators in the $osp(1|2)$
conformal field theory, there are several issues left for future works. \
For instance, we would like to connect this result with those presented in
the $osp(1|2)$ current algebra \cite{ERS, ER}. It should be interesting
since it appears in the quantization of two-dimensional supergravity in the
light-cone gauge \cite{PZ}. Moreover, the affine version of the $osp(1|2)$
algebra is the starting point in the construction of the $N=1$
superconformal minimal models by means of the Hamiltonian reduction
procedure \cite{BO}.

\vspace{.5cm}{}

{\bf Acknowledgment:} This work was supported in part by Funda\c{c}\~{a}o de
Amparo \`{a} Pesquisa do Estado de S\~{a}o Paulo--FAPESP--Brasil and by
Conselho Nacional de Desenvol\-{}vimento--CNPq--Brasil.

\end{document}